\documentclass[copyright,creativecommons]{eptcs} 
\providecommand{\event}{GandALF 2024} 

\usepackage[english]{babel}
\usepackage{iftex}
\usepackage{amssymb, amsmath}
\usepackage{amsthm}
\usepackage{csquotes}
\usepackage{graphicx}
\usepackage{subcaption}
\usepackage[ruled,vlined,linesnumbered]{algorithm2e}
\SetVlineSkip{0pt}   

\usepackage{enumitem}
\usepackage{tikz}
\usetikzlibrary{shapes.geometric}
\newcommand{\none}{\text -}
\newcommand{\Situation}[2]{$\underset{(#2)}{{#1}}$}
\tikzset{ ego/.style = {draw, circle, minimum size=0.8cm, inner sep=0pt}, alter/.style = {shape aspect=1,diamond,draw, minimum size=0.8cm, inner sep=-0.5pt}, winnable/.style = {fill=gray!50}}

\newcommand{\E}{EGO} 
\newcommand{\A}{ALTER}
\newcommand{\N}{\mathbb{N}}

\newtheorem{theorem}{Theorem}[section]

\newtheorem{definition}{Definition}[section]

\DeclareSymbolFont{largesymbolsA}{U}{txexa}{m}{n}
\DeclareMathSymbol{\varprod}{\mathop}{largesymbolsA}{16}

\ifpdf
  \usepackage{underscore}         
  \usepackage[T1]{fontenc}        
\else
  \usepackage{breakurl}           
\fi

\usepackage[capitalize, noabbrev, nameinlink]{cleveref} 

\title{Towards the Usage of Window Counting Constraints in the Synthesis of Reactive Systems to Reduce State Space Explosion}
\author{Linda Feeken
\institute{German Aerospace Center (DLR)\\ Oldenburg, Germany}
\email{linda.feeken@dlr.de}
\and
Martin Fr\"anzle
\institute{Carl von Ossietzky Universität Oldenburg\\ Oldenburg, Germany}
\email{fraenzle@informatik.uni-oldenburg.de}
}
\def\titlerunning{Window Counting Constraints for the Synthesis of Reactive Systems}
\def\authorrunning{L. Feeken \& M. Fr\"anzle}
\begin{document}
\maketitle

\begin{abstract}
The synthesis of reactive systems aims for the automated construction of strategies for systems that interact with their environment. 
Whereas the synthesis approach has the potential to change the development of reactive systems significantly due to the avoidance of manual implementation, it still suffers from a lack of efficient synthesis algorithms for many application scenarios. The translation of the system specification into an automaton that allows for strategy construction is nonelementary in the length of the specification in S1S and double exponential for LTL, raising the need of highly specialized algorithms. 
In this paper, we present an approach on how to reduce this state space explosion in the construction of this automaton by exploiting a monotony property of specifications. 
For this, we introduce window counting constraints that allow for step-wise refinement or abstraction of specifications.
In an iterating synthesis procedure, those window counting constraints are used to construct automata representing over- or under-approximations (depending on the counting constraint) of constraint-compliant behavior. Analysis results on winning regions of previous iterations are used to reduce the size of the next automaton, leading to an overall reduction of the state space explosion extend.
We present the implementation results of the iterated synthesis for a zero-sum game setting as proof of concept. Furthermore, we discuss the current limitations of the approach in a zero-sum setting and sketch future work in non-zero-sum settings.
\end{abstract}

\section{Introduction}

The automated translation of a system specification into its implementation is one of the most challenging problems in formal methods. Such a synthesis offers great potential in the development of new systems by significantly reducing the need for manual work in the engineering process.  
In this paper, we focus on synthesis for reactive systems, i.e. systems that are influenced by and interact with their environment. This interaction can be modeled as a game, in which the system tries to play according to its specification, whereas the moves of the environment can potentially impede the system from reaching its goal. Since the interaction between system and environment is typically of long-lasting nature without predefined end date, the game is infinite in the sense that a play of the game has infinite duration, while the arena, modeled as a graph, has finitely many states. The players play by moving a token from one state of the arena to the next. The player whose turn it is decides which of the outgoing transitions of the current state is chosen. A well-known type of game is the safety game: The system wins a play if it can avoid to reach predefined unsafe states. Otherwise, the environment wins. A player has a winning strategy, if it wins against all possible behavior of the other player. For two-player safety games on finite graphs, there always exists a winning strategy for one of the players and this winning strategy can be computed \cite{buchi1967solving}, \cite{thomas1995}. 
However, the efficient computation of winning strategies (not only in the case of safety games) is still an open challenge in the synthesis of reactive systems. 
A common synthesis approach is to generate a deterministic word automaton as game graph from specifications written as Linear Temporal Logic (LTL) formulae. Finally, a strategy that is winning in the game is calculated. By construction, the strategy automatically satisfies the specification. Unfortunately, the construction of the deterministic word automaton leads to an automaton with a number of states that is double-exponential in the length of the specification \cite{PnueliR89}, making the whole strategy synthesis unfeasible for many applications. For avoiding the most expensive part of the synthesis procedure, there exist synthesis algorithms that start with a subset of the specification language LTL, such that it is possible to construct the game graph in a more efficient way. One example for that is the usage of the LTL subclass Generalized Reactivity(1) (GR(1)), which allows to construct and solve the game in time $O(N^3)$ with $N$ being the size of the state space \cite{piterman2006synthesis}. While GR(1) is expressive enough for the specification of many systems \cite{maoz2015gr}, some specifications that do not fall into GR(1) remain unconsidered. For example, Maoz and Ringert mention the consideration of synthesis with counting patterns as future work in \cite{maoz2015gr}, but to the best of our knowledge, this is not yet done.

In this paper, we deal with the request for efficient synthesis for some types of counting patterns as part of the system specification and present the idea of iterated synthesis for such games. We call the considered counting patterns \enquote{window counting constraints}. These are of the form \\
\centerline{\textit{\enquote{The system plays action $act$ at most $k$ times out of $l$ of its own moves.}}}
with parameters $k, l\in \mathbb{N}, \, k\leq l$. The \enquote{at most} can also be replaced by \enquote{at least}. 
Such constraints arise naturally when the desired behavior of systems includes reoccurring elements. For example, an automated guided vehicle on a factory floor might need to charge its battery in at least two out of ten moves to avoid to get empty batteries on an exit path. The term \enquote{window} in the constraint type name  emphasizes the relation of those specifications to sliding windows in data stream monitoring \cite{PatroumpasS06}. For the sake of better readability, we also call them counting constraints in short.\\

We avoid the direct full translation of the specifications into a graph and instead focus on the following two observations:
(1) It is possible to influence how hard it is to satisfy a counting constraint by varying the parameters $k, l$ included in the counting constraints. More precisely, the (non-)existence of a strategy that fulfills the specification in a game with a set of counting constraints allows to make statements about the (non-)existence of such a strategy in a game with a set of counting constraints with varied parameters.
(2) The values in the counting constraints influence the scale of the game graph that encodes all information given by the constraints. The greater $k$ and $l$, the greater is the graph. Consequently, the values influence how much computational power and/or memory is needed in order to synthesize a winning strategy.

Combining these observations, the presented approach can be summarized as follows: 
Consider a two-player game graph and some specifications in the form of counting constraints. For solving the synthesis problem of finding a strategy for the system, such that the counting constraints are fulfilled,  
start with a subset of counting constraints that result in a small game graph or a trivially winnable game. Calculate winning strategies (if existent) and check what the (non-)existence of a winning strategy means for a game with refined/relaxed (depending on the constraints) constraints. 
This information shall give hints on which parts of the game with adapted values in the counting constraints are worth to investigate in the next iteration step and which parts of the game graph can then be neglected, leading to a reduction in the state space. In each iteration, the set of considered counting constraints converges more to the game of interest. 
Although the size of the game graphs may increase in each iteration, the gained state space reduction leads to a synthesis algorithm more efficient than when considering the game of interest as a whole from the beginning.
The motivation for starting with a game graph accompanied with counting constraints instead of a pure set of specifications comes from the robotic domain. In many applications, automated guided vehicles are moving in specified areas (like a factory floor). Modeling the setting as a game graph in which states encode the position of systems arises naturally.
However, the initial game graph can also represent the winning region of a priorly solved safety game  \cite{McNaughton93}, \cite{Zielonka98}, \cite{Jurdzinski00} that shall be accompanied with additional counting constraints. This way, it is possible to use the presented approach for safety games. Note that the safety game with neglected counting constraints is usually significantly smaller and hence easier to solve than the game with already included counting constraints.

This paper is structured as follows. 
In \cref{sec:relatedWork}, related work in the field of synthesis for reactive systems is presented, focusing on the challenge of constructing efficient algorithms.
After summarizing concepts and notations required to formulate the game, \cref{sec:gamesWithCountingConstraints} provides the definition of a game with counting constraints.
In \cref{sec:iteratedSynthesis}, we present the idea of iterated synthesis with counting constraints, including the results of a non-optimized implementation as proof-of-concept. The presented algorithm delivers promising results, but suffers of limitations that are targeted by our current research work. We discuss planned directions of future work in \cref{sec:discussion}. \Cref{sec:conclusion} concludes the paper.

\section{Related Work}\label{sec:relatedWork}

In 1957, Church formulated the Synthesis Problem as finding finite-memory procedures to transform an infinite sequence of input data into an infinite sequence of output data, such that the relation between input and output satisfies given specifications \cite{church1957}, \cite{Thomas09}. Around a decade later, Büchi and Landweber showed the decidability of the problem \cite{buchi1967solving}. However, the algorithmic complexity of synthesis algorithms remains a challenge. 
The translation of specifications from monadic second-order logic of one successor (S1S) into a Büchi automaton as part of the synthesis procedure is nonelementary in the length of specifications \cite{Stockmeyer74}. This indicates that it is not possible to construct a universally (or an in all cases) efficient synthesis algorithm that can handle complete S1S specifications.
For specifications expressible in Linear Temporal Logic (LTL), the problem is 2EXPTIME-complete \cite{PnueliR89}. \\
Acknowledging the absence of a generally low-complexity synthesis algorithm for arbitrary S1S/LTL specifications, the literature presents three primary approaches \cite{Finkbeiner2016}. 
(1) The first approach restricts the scope of considered specifications for synthesis to less expressive logics.
Here, the structure of the considered specifications is used to reduce the synthesis complexity.
(2) The second one is tackling the internal representation of the problem. Solutions following this approach are often aiming for algorithms with in average good runtime. In this approach, it suffices if most systems can be synthesized with acceptable resources (memory, computational time), while the existence of corner cases with worst-case complexity is accepted. 
(3) The third approach focuses on the output of the problem, the implementation. The size of the implementation is restricted, such that only small implementations are accepted as solutions of the synthesis problem. The rationale behind this is that small and hence less complicated implementations often exist for applications. Such solutions are often easier (that is, with less computational time) identifiable than bigger (complicated) implementations, if it is possible to steer the algorithm towards small solutions. Synthesis algorithms can follow more than one of those approaches.

A well studied class of specifications for approach (1) is General Reactivity of Rank 1 (GR1),  
a fragment of LTL for which there are symbolic synthesis algorithms that are polynomial in the size of the state space of the design \cite{piterman2006synthesis}.
Examples for other specification classes for which efficient solutions of the synthesis problem are investigated are Safety LTL \cite{ZhuTLPV17}, Metric Temporal Logic with a Bounded Horizon \cite{MalerNP07} and Extended Bounded Response LTL \cite{CimattiGGMT20}.

Following approach (2), Kupferman and Vardi developed a synthesis method that does not require the costly determinization of non-deterministic Büchi automata representing the specification \cite{KupfermanV05}, which is the most complex part in many synthesis algorithms. Other synthesis algorithms rely for instance on symbolic synthesis to represent sets of states of a game graph in a compact matter via antichains \cite{FiliotJR09}, \cite{FiliotJR11}, binary decision diagrams \cite{Ehlers10} and LTL fragements \cite{CimattiGGMT20}.

The work by Schewe and Finkbeiner presents a synthesis algorithm that employs bounded synthesis as approach (3). Their method uses translation of LTL specifications into sequences of safety tree automata, in order to constraint the size of the implementation \cite{ScheweF07a}.
\enquote{Lazy synthesis}, in which an SMT solver is used to construct potential implementations for an incomplete constraint system, extends the system only if required \cite{FinkbeinerJ12}.

The synthesis algorithm presented in this paper includes elements of approaches (2) and (3). We avoid the full construction of an automaton representing the specifications by starting with a small specification that is successively enlarged. In each step, the size of the resulting automaton is reduced (if possible). The procedure stops, if a winning strategy can already be found in some intermediate step, leading to small solutions. However, it is not possible to restrict the size of the implementation directly as commonly done in bounded synthesis. 

The general idea is inspired by the work of Chen et al. on games with delay. In this work, one player only receives information on the moves of the environment with a delay of $k \in \mathbb{N}$ turns. For strategy synthesis, the delay is incrementally enlarged from zero to $k$ with a graph reduction step after each iteration step \cite{chen2018s}.

\section{Games with Counting Constraints} \label{sec:gamesWithCountingConstraints}
This section introduces games with counting constraints after repeating standard definitions for two-player games that are needed to formalize the presented game. 

\begin{definition}[Two-player game graph]
	A \textbf{two-player finite-state game graph} is of the form $G = (S, s_0, S_{\E}, S_{\A}, \Sigma_{\E}, \Sigma_{\A}, \rightarrow)$ where $S$ is a finite (non-empty) set of states, $S_{\E}$, $S_{\A}$ define a partition of $S$, $s_0 \in S_{\E}$ is the initial state, $\Sigma_{\E}$ is a finite alphabet of actions for player $\E$,  $\Sigma_{\A}$ is a finite alphabet of actions for player $\A$ and $\rightarrow \subseteq S \times (\Sigma_{\E} \cup \Sigma_{\A}) \times S$ is a set of labeled transitions satisfying the following four conditions:
	\begin{itemize}
		\item Bipartition: For each $(s,\sigma, s') \in \rightarrow$ holds either (1) $s\in S_{\E}$ and $s' \in S_{\A}$ or (2) $s \in S_{\A}$ and $s' \in S_{\E}$.
		\item Absence of deadlock: For each $s \in S$ there exists $\sigma \in \Sigma_{\E} \cup \Sigma_{\A}$ and $s' \in S$, such that $(s,\sigma,s') \in \rightarrow$.
		\item Alphabet restriction on actions: For a player $p \in \{\E, \A\}$ holds: If $(s,\sigma,s') \in \rightarrow$ with $s \in S_{p}$, then $\sigma \in \Sigma_{p}$.
		\item Determinacy of moves: For $p \in \{\E, \A\}$ and $\sigma \in \Sigma_{\E} \cup \Sigma_{\A}$ holds: if $s \in S_{p}$ and $(s,\sigma,s'), (s,\sigma,s'') \in \rightarrow$, then $s' = s''$.  
	\end{itemize}
\end{definition}

Such a game graph, also referred to as \enquote{arena}, encodes a game between the two players $\E$ and $\A$. For $p \in \{\E, \A\}$ the set of states $S_p$ contains the states where it is the turn of player $p$ to perform an action, also called \enquote{$p$ controls $s$}. Due to the bipartition and alphabetic restriction on actions, the game is \enquote{turn-based}, i.e.\ the two players alternate between choosing one of the possible actions. Since the game graph does not contain deadlocks, it results in an infinite sequence of states and actions, called an infinite play.

\begin{definition}[Infinite play]
	Let  $G = (S, s_0, S_{\E}, S_{\A}, \Sigma_{\E}, \Sigma_{\A}, \rightarrow)$ be a two-player game graph. An \textbf{infinite play} on $G$ is an infinite sequence $\pi = (\pi_i \sigma_i)_{i\in \N_0} = \pi_0 \sigma_0 \pi_1 \sigma_1 \dots$ with $\pi_0 = s_0$ and $\pi_i \sigma_i \pi_{i+1} \in \rightarrow$ for all $i \in \N_0$. $\Pi(G)$ denotes the set of all infinite plays on $G$. 
\end{definition}

In such an infinite play, the two players play against (or in case of collaborative games: with) each other. Players can have strategies that determine how they react in each step of the play.

\begin{definition}[Strategy]
	Let  $G = (S, s_0, S_{\E}, S_{\A}, \Sigma_{\E}, \Sigma_{\A}, \rightarrow)$ be a two-player game graph.
	\begin{itemize}
		\item For a play $\pi = (\pi_i \sigma_i)_{i\in \N_0}$, a \textbf{prefix} of $\pi$ up to position $n$ is denoted by $\pi(n) = \pi_0 \sigma_0 \pi_1 \dots \pi_{n-1} \sigma_{n-1} \pi_n$. The length of $\pi(n)$, denoted by $|\pi(n)|$, is $n+1$. The last state $\pi_n$ of $\pi(n)$ is called the \textbf{tail} of the prefix $\pi(n)$, denoted by $Tail(\pi(n))$. The set of all prefixes of plays in the game graph $G$ is $Pref(G)$.
		\item For a player $p \in \{\E, \A\}$ and a game graph $G$, the set of all prefixes that end in a state controlled by $p$ is $Pref_p(G) := \{\pi(n) \in Pref(G) \, | \, Tail(\pi(n)) \in S_p\}$.
		\item A \textbf{strategy} for a player $p \in \{\E, \A\}$ in the game graph $G$ is a mapping $\varphi: \, Pref_{p}(G) \longrightarrow 2^{\Sigma_{p}}$, such that for all prefixes $\pi(n) \in Pref_p(G)$ and all $\sigma \in \varphi(\pi(n))$ there exist a state $s \in S \setminus S_p$ and a transition $(Tail(\pi(n)), \sigma, s) \in \rightarrow$. 
		\item The \textbf{outcome} $O(G, \varphi)$ of a strategy $\varphi$ of  $p \in \{\E, \A\}$ in the game graph $G$ is the set of all possible plays when player $p$ follows the strategy $\varphi$ and the other player plays arbitrary, i.e.\ $O(G, \varphi) := \{ \pi = (\pi_i \sigma_i)_{i\in \N_0} \in \Pi(G) \, | \, \forall i \in \N_0: \sigma_{2i} \in \varphi(\pi(2i)) \text{ if } s_0 \in S_p \text{ and } \sigma_{2i+1} \in \varphi(\pi(2i+1)) \text{ otherwise}\}$.
	\end{itemize}
\end{definition}

In a safety game, the player $\E$ wins, if it has a strategy that guarantees to never visit predefined unsafe states. The environment, on the other hand, wins if an unsafe state is reached. Hence, each play of a two-player safety game always has exactly one winner and one loser. Games with this property are called zero sum games.

\begin{definition}[Safety Game]
	A \textbf{safety game} $G = (G',\mathcal{U})$
	consists of a two-player finite-state game graph $G' = (S, s_0, S_{\E}, S_{\A}, \Sigma_{\E}, \Sigma_{\A}, \rightarrow)$ and a set $\mathcal{U} \subseteq S$ of unsafe states.\\
	Player $\E$ has a \textbf{winning strategy} $\varphi$ on $G$, if $\varphi$ is a strategy on $G'$, such that none of the plays in $O(G', \varphi)$ include a state $u\in \mathcal{U}$.
	The \textbf{winning region} of $G$ is defined as the set of states $\tilde{S}\subseteq S$, where $\E$ can win from any state $s\in \tilde{S}$. 
	This means $\E$ has a winning strategy in the game $\tilde{G}_s$ with $\tilde{G}_s = (S, s, S_{\E}, S_{\A}, \Sigma_{\E}, \Sigma_{\A}, \rightarrow, \mathcal{U})$.
\end{definition}

We are now introducing window counting constraints as a mean to encode reoccurring behavior of the player $\E$ with limits on which action can be selected how often in each snippet (or: window) of a play of a given length.
\begin{definition}[Window Counting Constraints]\label{def:windowCountingConstraints}
	Let $G$ be a game graph with the two players $EGO$ and $ALTER$, denote with $a$ an action and $k, l \in \mathbb{N}$ with $k \leq l$. Let $\pi = (\pi_i \sigma_i)_{i\in \mathbb{N}_0}$ be a play on $G$.
	\begin{enumerate}
		\item $\textbf{CC_{max}(\E, a, k, l)}$ is defined as the abbreviation for \enquote{The player $\E$ plays action $a$ at most $k$ times out of $l$ of its own turns.}\\ 
		$CC_{max}(\E, a, k, l)$ is satisfied on $\pi$, if for all $i\in \mathbb{N}_0$ holds $|\{\sigma_{2m} \, | \, \sigma_{2m}=a,\, i\leq m \leq i+l\}|\leq k$.
		\item $\textbf{CC_{min}(EGO, a, k, l)}$ is defined as the abbreviation for \enquote{The player $\E$ plays action $a$ at least $k$ times out of $l$ of its own turns.}\\  
		$CC_{min}(\E, a, k, l)$ is satisfied on $\pi$, if for all $i\in \mathbb{N}_0$ holds $|\{\sigma_{2m} \, | \, \sigma_{2m}=a,\, i\leq m \leq i+l\}|\geq k$.		
	\end{enumerate}
	A prefix of a play on $G$ satisfies a counting constraint, if it can be complemented to an infinite play that satisfies the counting constraint in any way (in particular, the extended prefix does not need to be a play on $G$). 
	The parameter $l$ is called the length of a counting constraint.
\end{definition}

The above definition might raise the question why we do not consider similar counting constraints for the player $\A$, representing the environment. Such constraints impose a set of challenges, which we will discuss in \cref{sec:discussion} and plan to tackle as future work. 

We extend the definition of satisfying a counting constraint for a play canonically to satisfying a set of counting constraints and counting constraints being satisfied on a strategy. \\
In a (zero-sum) game with counting constraints, the $\E$ player needs to satisfy all of its counting constraints in order to win the game.

\begin{definition}[Games with Counting Constraints]\label{def:gameWithCounstraints}
	A two-player \textbf{game with counting constraints} is defined as  $G = (G', CC_{\E})$, where 
	\begin{itemize}
		\item $G' = (S, s_0, S_{\E}, S_{\A}, \Sigma_{\E}, \Sigma_{\A}, \rightarrow)$ is a two-player finite-state game graph.
		\item $CC_{\E} \subset \{CC_{m}(\E, a, k, l) \, | \, m\in\{min, max\}, k,l \in \mathbb{N}, k\leq l, a \in \Sigma_{\E}\}$ is a finite sets of counting constraints of $\E$
	\end{itemize}
	Player $\E$ wins a play on $G$, if the play satisfies all counting constraints $CC_{\E}$. Otherwise, $\A$ wins.
	A strategy $\varphi$ of $\E$ is winning for $\E$ (or a \textbf{winning strategy} of $\E$), if $\E$ wins all plays in $O(G, \varphi)$.
\end{definition}

\section{Iterated Synthesis with Counting Constraints}\label{sec:iteratedSynthesis}

The key advantage of counting constraints for synthesis is their monotony property: If $\E$ has a strategy, such that $\E$ plays an action $a$ at most $k$ times in $l$ turns (i.e.\ the strategy satisfies $CC_{max}(\E, a, k, \allowbreak l)$), then $\E$ also plays $a$ at most $k$ times in $l-1$ turns (i.e. the strategy satisfies $CC_{max}(\E, a, k, l-1)$).  
In other words: The existence of a winning strategy for a game with counting constraint $CC_{max}(\E, a, k, \allowbreak l-1)$ is a necessary condition for the winning strategy for a game with $CC_{max}(\E, a, k, l)$. Moreover, only a strategy that fulfills $CC_{max}(\E, a, k, l-1)$ can also fulfill $CC_{max}(\E, a, k, l)$.
From an algorithmic perspective, it is more favorable to search for strategies that satisfy $CC_{max}(\E, a, k, l-1)$ then for strategies that satisfy $CC_{max}(\E, a, k, l)$, since the graph that encodes the first (shorter) constraint is smaller than the one that encodes the latter (longer) constraint. Intuitively, this is caused by more memory that is needed for remembering the last $l$ own turns instead of only $l-1$ turns. The synthesis idea is related to the incremental approach used by synthesis with antichains \cite{FiliotJR09}.\\
For a counting constraint of the form  $CC_{min}(\E, a, k, l-1)$ (\enquote{$\E$ plays action $a$ at least $k$ times out of $l-1$ of its turns}), we can conduct that if a strategy fulfills the constraint, it automatically also fulfills the larger constraint  $CC_{min}(\E, a, k, l)$. Hence, if we already have a strategy that fulfills $CC_{min}(\E, a, k, l-1)$, it is needless to do the more challenging search for a strategy that fulfills $CC_{min}(\E, a, k, l)$.  
\begin{theorem}\label{theorem:monotony}
	Let  $G = (S, s_0, S_{\E}, S_{\A}, \Sigma_{\E}, \Sigma_{\A}, \rightarrow, CC_{\E})$ be a two-player game with counting constraints. 
	\begin{enumerate}
		\item For $CC_{max}(EGO, a, k, l) \in CC_{\E}$ holds: If $\varphi$ is a winning strategy for $\E$ on $G$, then it is also a winning strategy for $\E$ on $G'$, where $G'$ equals $G$ except that $CC_{max}(EGO, a, k, l)$ is exchanged by $CC_{max}(EGO, a, k, l-1)$.
		\item For $CC_{min}(EGO, a, k, l) \in CC_{\E}$ holds: If $\varphi$ is a winning strategy for $\E$ on $G$, then it is also a winning strategy for $\E$ on $G'$, where $G'$ equals $G$ except that $CC_{max}(EGO, a, k, l)$ is exchanged by $CC_{max}(EGO, a, k, l+1)$.
	\end{enumerate}
\end{theorem}
Since the proof is straightforward, we omit it here.
It is also possible to vary the $k$ parameter in the constraints instead of $l$ with similar conclusions. 
With each of those iterations, the number of previously made turns that need to be memorized is increasing. We introduce situation graphs as a mean to encode the relevant history of a play into game graphs. In a nutshell, a situation is a state of the game graph $G$ combined with the counting constraint-relevant part of the history on how the state was reached. It allows for categorizing states of the game into \enquote{part of the winning region} and \enquote{not winnable}, which reduces a game with counting constraints to a classical safety game with states in which $\E$ violates its constraints as unsafe states. 

\begin{definition}[Situation Graph]
	Let $G = (S, s_0, S_{\E}, S_{\A}, \Sigma_{\E}, \Sigma_{\A}, \rightarrow, CC_{\E})$ be a game with counting constraints. Fix some order $CC_{\E} = \{C_{\E,1},\dots, C_{\E,q}\}$ of the counting constraints.
	For each counting constraint $C=CC_m(\E,a,k,l) \in CC_{\E}$, $m\in \{min, max\}$ define a transition 
	$$h_C\colon \{0,1,\text{none}\}^l \times \Sigma_{\E} \to \{0,1,\text{none}\}^l, \quad 
	((v_1,\dots,v_l), act) \mapsto \biggl\{
	\begin{aligned}
		(1,v_1,\dots,v_{l-1}), &\text{ if } act=a\\
		(0,v_1,\dots,v_{l-1}), &\text{ else.}
	\end{aligned} $$
	A situation is a tuple $(s, H_{\E})$ with $s\in S$ being a state in $G$, $H_{\E} \in \varprod_{i=1}^q \text{codom}(h_{C_{\E,i}})$ and $\text{codom}(f) \allowbreak =Y$ denoting the codomain of a function $f\colon X\rightarrow Y$.  
	Denote the set of all situations by $\tilde{S}$.
	Define a transition 
	\begin{align*}
		\hookrightarrow' \colon \tilde{S} \times \Sigma_{\E} &\to \tilde{S} \\ ((s, (v_1,\dots,v_q)),act) &\mapsto 
		\biggl\{
		\begin{aligned}
			(s',(h_{C_{\E,1}}(v_1, act),\dots,h_{C_{\E,q}}(v_q, act)), &\text{ if } s\in S_{\E}\\
		    (s', (v_1,\dots,v_q)), &\text{  if } s\in S_{\A}
		\end{aligned} 
	\end{align*}
	such that $(s,act,s')\in \rightarrow$. 
	The transition $\hookrightarrow'$ defines how to get from one situation to another when using the transition $\rightarrow$ in $G$.\\
	A situation $(s, H_{\E})$ is satisfying a counting constraint $CC_{min}(\E,a,k,l) \in CC_{\E}$, if for the corresponding part $(v_1,\dots,v_l)$ in $H_{\E}$ holds $|\{v_i \, | v_i=1, i=1,\dots,l\}|\leq k$. Similarly, the situation satisfies $CC_{max}(\E,a,k,l) \in CC_{\E}$, if $|\{v_i \, | v_i=1, i=1,\dots,l\}|\geq k$. \\
	The situation graph of $G$ is the two-player finite game graph $Sit = (S', s_{init},S'_{\E}, S'_{\A}, \Sigma_{\E}, \Sigma_{\A}, \allowbreak \hookrightarrow)$ with
		\begin{itemize}
		\item initial state being the situation $s_{init}=(s_0, H_{init, \E})$ with all entries in $H_{init, \E}$ being $none$,
		\item transition relation $\hookrightarrow \subseteq \tilde{S} \times \Sigma_{\E} \times \tilde{S}$ with $(s,act,s')\in \hookrightarrow$ 
		\item set of states $S'$ being all situations that are reachable from $s_{init}$ via $\hookrightarrow$,
		\item $S'_{p}\subseteq S'$ the states $(s,H_{\E})$ that are controlled by player $p\in \{\E,\A\}$, that is $s\in S_p$.
	\end{itemize}
	The winning region of $\E$ in the situation graph is the set of states $\tilde{S}\subseteq S'$ from which $\E$ has a winning strategy, that is, from which $\E$ can guarantee to only visit states that satisfy all counting constraints in $CC_{\E}$.
\end{definition}

The situation graph of a game is a deterministic Büchi automaton that represents the full specification of $\E$, if the complete set of counting constraints is considered. Counting constraints are expressible as (long) LTL-formulae, hence, using the full situation graph for synthesis is generally only doable in time double-exponential in the size of the LTL-specification \cite{PnueliR89}. 
The iterated synthesis approach avoids to construct the full situation graph. The general idea is to start with a rather small game by using counting constraints of small lengths and iterate over the length. In each iteration, a part of the corresponding situation graph is constructed and analyzed and knowledge that can be reused in following iterations is identified. This knowledge is determining which parts of the situation graph for the next iteration needs to be constructed and which parts can be omitted, relying on \cref{theorem:monotony}.

For iteration over one counting constraint $CC_{min}(\E, a, k, l)$, the synthesis procedure is sketched in \cref{alg:iteratedSynthesis} and algorithms called therein. For better readability, the algorithms only handle one other counting constraint $CC_{max}(\E, b, m, n)$ besides the one that is iterated over. However, since the other counting constraint remains fixed during the iteration approach, it is possible to add additional (fixed) counting constraints with only minor adaptions.
\Cref{alg:iteratedSynthesis} basically alternates between calling two other algorithms: Starting with the smallest possible counting constraint $CC_{min}(\E, a, k, k)$, the situation graph for the respective game is generated (\cref{alg:generateSituationGraph}). After that, the resulting graph is analyzed in order to find the winning region for $\E$ (\cref{alg:findWinReg}). If the initial state of the situation graph belongs to the winning region, a set of winning strategies for $\E$ is found and the algorithm terminates. If the initial state is not winnable, the next iteration starts with the next longer counting constraint. If even the winning region of the situation graph for $CC_{min}(\E, a, k, l)$ does not contain the initial state of the situation graph, no winning strategy for $\E$ exists. 

\cref{alg:generateSituationGraph} generates (parts of) the situation graphs in each iteration. States of the situation graph are called \enquote{situations} in the algorithm in order to avoid confusion with the states of the underlying game graph. Note that the algorithm omits successors of states that violate counting constraints of $\E$, since those states do not belong to the winning region (line \ref{alg_line:situationsDontViolateECoCocs}). 
In the first iteration, there is no additional information on winnable states available, hence the full situation graph (minus successors of states violating constraints of $\E$) needs to be constructed. Due to the small counting constraint length, this graph is significantly smaller than it would be for the full constraint length. As soon as \cref{alg:findWinReg} identifies any winnable states, this knowledge can than be used in the construction of the situation graph in the next iteration: The construction begins with adding the initial state to an empty (directed) graph. Successors of already added states are added successively. For each added state, it is checked if there is a \enquote{related} state in the winning region of the previous iteration. If this is the case, the state can also be marked as being in the winning region and successors do not need to be considered. As a consequence, the situation graph is only partly constructed, saving computational time and memory.
A state $s$ of a situation graph in one iteration for a counting constraint with action $a$ is related to a state $s'$ of the situation graph of the previous iteration, if $s$ can be transformed into $s'$ by only deleting the last entry of the history of $a$. The identification of such states is the key factor for more efficient synthesis via the presented approach, since it allows to perform synthesis on incomplete graphs, allowing for a pruning step in each iteration.

\cref{alg:findWinReg} calculates the winning region for a given (incomplete) situation graph. The reduction of the size of the graph by incomplete construction is again speeding up the algorithm. States of the situation graph without successor are considered first. Such states are either already identified as being winnable since they are related to winnable states of the previous iteration (line 1) or can be marked as non-winnable (aka $losing$, line 2), since the counting constraint of $\E$ is violated. 
The rest of the algorithm is rather generic and uses a version of fixed point computation for a finite-state two-player safety game with the already identified states in $losing$ as unsafe states.\\

In iterations over counting constraints of the form $CC_{max}(EGO, a, k, l)$, it is searched for states of the situation graph that are not in the winning region of $\E$. Such states will also not be visited by winning strategies in the following iterations. Except for searching for non-winnable states instead of winnable states, the synthesis procedure is similar to the one for $CC_{min}(EGO, a, k, l)$ constraints. If the initial state of a situation graph in any iteration is marked as non-winnable, there exists no winning strategy for $\E$. 

\begin{algorithm}
	\scriptsize
	\DontPrintSemicolon
	\KwIn{$G = (S, s_0, S_{\E}, S_{\A}, \Sigma_{\E}, \Sigma_{\A}, \rightarrow, \{CC_{min}(\E, a,k,l), CC_{max}(\E, b,m,n)\})$ - two-player game with counting constraints.}
	\KwOut{If winning strategy for $\E$ exists: \\
		\, $situation\_graph$ - part of the smallest situation graph in which a winning strategy exists \\ 
		\, $winning\_situations$ - set of states of the situation graph, forming a subset of the winning region for the graph \\
		Else: LOSING - no winning strategy for $\E$ exists}
	$winning\_region\longleftarrow$ empty directed graph\;
	\For(\tcp*[f]{increase $\E$ counting constraint length from $k$ to $l$}){$c \in \{k, \dots,l\}$}{
		$situation\_graph$, $\longleftarrow$ generate\_situation\_graph($G$,$k$,$c$,$m$,$n$,$winning\_region.states$) \;
		$winning\_region \longleftarrow$ find\_winning\_region($situation\_graph$, \{$CC_{min}(\E, a,k,c)$, $CC_{max}(\E, b,m,n)\}$, $S_{\E}$, $S_{\A}$)\;
		\If{$situation\_graph.initial\_situation \in winning\_region.states$}{
		\Return{$situation\_graph$, $winning\_region$}
		}
	}
	\Return{LOSING}
	\caption{Iterated Synthesis over one $CC_{min}(\E, a,k,l)$ counting constraint \label{alg:iteratedSynthesis}}
\end{algorithm}

\begin{algorithm}
	\SetAlgoSkip{}
	\scriptsize
	\DontPrintSemicolon
	\KwIn{$G = (S, s_0, S_{\E}, S_{\A}, \Sigma_{\E}, \Sigma_{\A}, \rightarrow, \{CC_{min}(\E, a,k,c), CC_{max}(\E, b,m,n)\})$  two-player safety game with counting constraints; $previous\_winning\_situations$ set of winnable states of the situation graph for $G$ with $CC_{min}(\E, a,k,c-1)$}
	\KwOut{$situation\_graph$ - situation graph of $G$ without unfolding areas that are already winnable in the previous iteration (considering $CC_{min}(\E, a,k,c-1)$)}
	$initial\_situation$ $\longleftarrow$ ($s_0$,[\text{none for $i$ in range($c$)}], [\text{none for $i$ in range($n$)}]);
	$unfinished\_situations$ $\longleftarrow$ $\{initial\_situation\}$\;
	$finished\_situations$ $\longleftarrow$ $\emptyset$;
	$winning\_situations$ $\longleftarrow$ $\emptyset$\;
	$situation\_graph \longleftarrow$ empty directed graph; $situation\_graph.situations$ $\longleftarrow$ $\{initial\_situation\}$\;
	$A \longleftarrow CC_{min}(\E, a,k,c)$; $B \longleftarrow CC_{max}(\E, b,m,n)$\;
	\While(\tcp*[f]{while not all successors of states in the graph are considered}){$unfinished\_situations$}{
		\tcc{take a situation from $unfinished\_situations$ and add all needed successors to the graph}
		choose any $current\_situation \in unfinished\_situations$; 
		$unfinished\_situations$.remove($current\_situation$)\;
		$all\_next\_moves \longleftarrow \{ (current\_situation.state,act,s') \in (S_{\E}\cup S_{\A}) \times (\Sigma_{\E} \cup \Sigma_{\A}) \times (S_{\E}\cup S_{\A}) \, | \, (current\_situation.state, act, s') \in \rightarrow\}$ \;
		\For{$next\_move \in all\_next\_moves$}{
			\If(\tcp*[f]{case: $\E$ controls the current situation}){$current\_situation.state \in S_{\E}$}{
				\tcc{construct one successor of $current\_situation$ in the situation graph}
				$next\_situation \longleftarrow (next\_move.tail, [next\_move.action == a, current\_situation.history_{\E.A}[:-1]], [next\_move.action == a,current\_situation.history_{\E.B}[:-1]])$\;
				\If(\tcp*[f]{case: situation not yet in the situation graph}){$next\_situation \notin situations$}{
					$situation\_graph.situations$.add($next\_situation$)\;
					\lIf{$next\_situation$ does not satisfy $A$ or $B$ \label{alg_line:situationsDontViolateECoCocs}}{
						$finished\_situations$.add($next\_situation$)		
					}	
					\Else(\tcp*[f]{check if $next\_situation$ is related to a winnable situation of prev. iteration}){
						$related\_next\_situation \longleftarrow (next\_situation.state, next\_situation.history_{\E.A}[:-1], next\_situation.history_{\E.B})$\;
						\If{$related\_next\_situation \in previous\_safe\_situations$}{
							$winning\_situations$.add($next\_situation$);
							$finished\_situations$.add($next\_situation$)
						}
						\Else{
							\lIf{$next\_situation \notin finished\_situations$}{$unfinished\_situations$.add($next\_situation$)}
						}
					}
				}
				$situation\_graph.transitions$.add($(current\_situation, next\_move.action, next\_situation)$)
			}
			\Else(\tcp*[f]{case: $\A$ controls the current situation}){
				$next\_situation \longleftarrow (next\_move.tail, current\_situation.history_{\E})$\;
				\If{$next\_situation \notin situations$}{
					$situation\_graph.situations$.add($next\_situation$)\;
					$related\_next\_situation \longleftarrow (next\_situation.state, next\_situation.history_{\E.A}[:-1], next\_situation.history_{\E.B})$\;
					\If{$related\_next\_situation \in previous\_safe\_situations$}{
						$winning\_situations$.add($next\_situation$); 
						$finished\_situations$.add($next\_situation$)
					}
				}
				$situation\_graph.transitions$.add($(current\_situation, next\_move.action, next\_situation)$)\;
				\lIf{$next\_situation \notin finished\_situations$}{
					$unfinsihed\_situations$.add($next\_situation$)
				}
			}						
		}
		$finished\_situations$.add($current\_situation$)
	}
	\Return{$situation\_graph$}
	\caption{\textbf{generate\_situation\_graph}: Construction of the situation graph without unfolding regions already won in previous iterations} \label{alg:generateSituationGraph}
\end{algorithm}

\begin{algorithm}
	\scriptsize
	\DontPrintSemicolon
	\KwIn{$situation\_graph$ as constructed in \cref{alg:generateSituationGraph}; 
	$CC_{min}(\E, a, k, c)$, $CC_{max}(\E, b, m, n)$ counting constraints belonging to $situation\_graph$; 
 	$S_{\E}$, $S_{\A}$ states of the underlaying game}
	\KwOut{part of the winning region for $\E$ in $situation\_graph$}
	\tcc{Divide states of $situation\_graph$ without successor into winnable and losing states}
	$winning \longleftarrow \{ sit \, | \, \text{state } sit \text{ has no successor and satisfies } CC_{min}(\E, a, k, c) \text{ and } CC_{max}(\E, b, m, n)\}$\;
	$losing \longleftarrow \{ sit \, | \, \text{state } sit \text{ has no successor and does not satisfy } CC_{min}(\E, a, k, c) \text{ or } CC_{max}(\E, b, m, n)\}$\;
	\tcc{mark predecessors of winning $\A$-situations as winning}
	$winning$.add($\{pred \, | \, pred \text{ is predecessor of some } sit\in winning \text{ with }  sit.state\in S_{\E}\}$)\;
	\tcc{mark $\A$-situations as winning, if all successors are winning}
	$winning$.add($\{sit \, | \, sit.state \in S_{\A}, \text{for all successors } suc \text{ of } sit \text{ holds: } suc \in winning\}$)\;
	\tcc{identify losing states}
	$progress \leftarrow$ TRUE\;
	\While{$progress$}{
		$progress \leftarrow$ FALSE\;
		\tcc{handle all situations controlled by $\E$ and marked as losing}
		$losing\_\E\_sit = \{situation \, | \, situation.state\in S_{\E}\} \cap losing$\;
		\If{$losing\_\E\_sit$}{
			$losing$.add($\{predecessor \, | \, \exists sit \in losing\_\E\_sit: sit \text{ is a successor of } predecessor \}$)\;
			\tcc{delete all ingoing and outgoing transitions from states in $losing\_\E\_sit$ and those states itself from $situation\_graph$}
			$situation\_graph$.remove\_nodes\_from($losing\_\E\_sit$); $progress \longleftarrow$ TRUE
		}
		\tcc{handle situations controlled by $\A$ \& already marked as losing}
		$losing\_\A\_sit \leftarrow \{situation \, | \, situation.state\in S_{\A}\} \cap losing$\;
		\If{$losing\_\A\_sit$}{
			\tcc{delete all ingoing and outgoing transitions from states in $losing\_\A\_sit$ and those states itself from $situation\_graph$}
			$situation\_graph$.remove\_nodes\_from($losing\_\A\_sit$); $progress \longleftarrow$ TRUE
		}
		\tcc{handle situations not marked as winning and without successor}
		$no\_win \leftarrow \{situation \, | \, situation\notin winning, situation \text{ has no successor in } situation\_graph\}$\;
		\lIf{no\_win}{
			$losing$.add($no\_win$); $progress \longleftarrow$ TRUE
		}
	}
	\Return{$situation\_graph$}
	\caption{\textbf{find\_winning\_region}} \label{alg:findWinReg}
\end{algorithm}

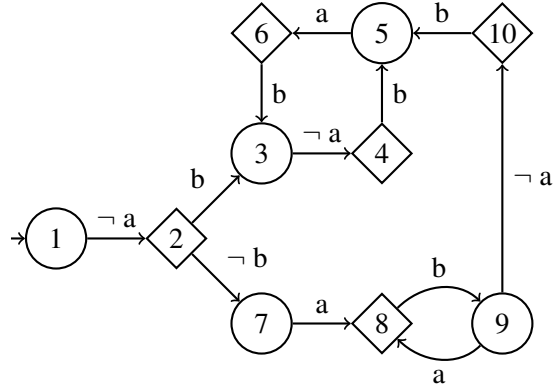
\begin{figure}
	\centering
	\begin{tikzpicture}[node distance={16mm}, thick, scale=.6] 
		\node[ego] (1) {$1$}; 
		\node[alter] (2) [right of=1] {$2$}; 
		\node[ego] (3) [above right of=2] {$3$};
		\node[alter] (4) [right of=3] {$4$};
		\node[ego] (5) [above of=4] {$5$};
		\node[alter] (6) [above of=3] {$6$};
		\node[ego] (7) [below right of=2] {$7$};
		\node[alter] (8) [right of=7] {$8$};
		\node[ego] (9) [right of=8] {$9$};
		\node[alter] (10) [right of=5] {$10$};
		
		\draw[->] (-1,0) -- (1); 
		\draw[->] (1) -- node[midway, above] {$\neg$ a} (2); 
		\draw[->] (2) -- node[midway, above left] {b} (3); 
		\draw[->] (3) -- node[midway, above] {$\neg$ a} (4);
		\draw[->] (4) -- node[midway, right] {b} (5); 
		\draw[->] (5) -- node[midway, above] {a} (6); 
		\draw[->] (6) -- node[midway, right] {b} (3);
		\draw[->] (2) -- node[midway, above right] {$\neg$ b} (7); 
		\draw[->] (7) -- node[midway, above] {a} (8); 
		\draw[->] (8) to [out=45,in=135]  node[midway, above] {b} (9); 
		\draw[->] (9) to [out=225,in=315]  node[midway, below] {a} (8); 
		\draw[->] (9) -- node[midway, right] {$\neg$ a} (10); 
		\draw[->] (10) -- node[midway, above] {b} (5); 
	\end{tikzpicture} 
	\caption{Two-player game graph. States represented as circles are controlled by $\E$, diamond-shaped states are controlled by $\A$. $\E$ shall fulfill the counting constraint $CC_{min}(\E, a,1,7)$ ($\E$ plays $a$ at least one time in 7 turns).}
	\label{fig:smallGameIterationsMatter}
\end{figure}

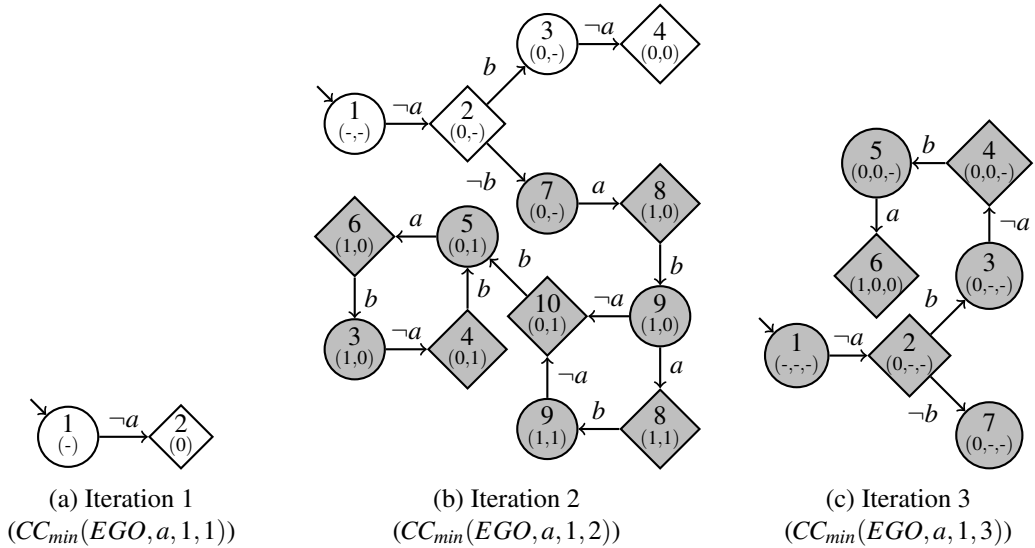
\begin{figure}
	\centering
	\small
	\begin{subfigure}[b]{.32\textwidth}
		\centering
		\begin{tikzpicture}[node distance={15mm}, thick, scale=1.0, font=\small] 
			\node[ego] (1) {\Situation{1}{\none}}; 
			\node[alter] (2) [right of=1] {\Situation{2}{0}}; 
			
			\draw[->] (-0.5,0.5) -- (1); 
			\draw[->] (1) -- node[midway, above] {$\neg a$} (2); 
		\end{tikzpicture}   
		\caption{Iteration 1 \\ ($CC_{min}(\E, a,1,1)$)}
		\label{fig:iteration1}                                     
	\end{subfigure}%
	\begin{subfigure}[b]{.32\textwidth}
		\centering
		\begin{tikzpicture}[node distance={15mm}, thick, scale=1.0, font=\small, align=center] 
			\node[ego] (12) {\Situation{1}{\none,\none}}; 
			\node[alter] (22) [right of=12] {\Situation{2}{0,\none}}; 
			\node[ego] (321) [above right of=22] {\Situation{3}{0,\none}};
			\node[ego, winnable] (72) [below right of=22] {\Situation{7}{0,\none}};
			\node[alter] (421) [right of=321] {\Situation{4}{0,0}};
			\node[alter, winnable] (821) [right of=72] {\Situation{8}{1,0}};
			\node[ego, winnable] (921) [below of=821] {\Situation{9}{1,0}};
			\node[alter, winnable] (822) [below of=921] {\Situation{8}{1,1}};
			\node[ego, winnable] (922) [left of=822] {\Situation{9}{1,1}};
			\node[alter, winnable] (102) [left of=921] {\Situation{10}{0,1}};
			\node[ego,winnable] (52) [above left of=102] {\Situation{5}{0,1}};
			\node[alter, winnable] (62) [left of=52] {\Situation{6}{1,0}};
			\node[ego, winnable] (322) [below of=62] {\Situation{3}{1,0}};
			\node[alter, winnable] (422) [right of=322] {\Situation{4}{0,1}};
			
			\draw[->] (-0.5,0.5) -- (12); 
			\draw[->] (12) -- node[midway, above] {$\neg a$} (22); 
			\draw[->] (22) -- node[midway, above left] {$b$} (321); 
			\draw[->] (22) -- node[midway, below left] {$\neg b$} (72);
			\draw[->] (321) -- node[midway, above] {$\neg a$} (421); 
			\draw[->] (72) -- node[midway, above] {$a$} (821);
			\draw[->] (821) -- node[midway, right] {$b$} (921);
			\draw[->] (921) -- node[midway, right] {$a$} (822);
			\draw[->] (822) -- node[midway, above] {$b$} (922);
			\draw[->] (922) -- node[midway, right] {$\neg a$} (102);
			\draw[->] (921) -- node[midway, above] {$\neg a$} (102);
			\draw[->] (102) -- node[midway, above right] {$b$} (52);
			\draw[->] (52) -- node[midway, above] {$a$} (62);
			\draw[->] (62) -- node[midway, right] {$b$} (322);
			\draw[->] (322) -- node[midway, above] {$\neg a$} (422);
			\draw[->] (422) -- node[midway, right] {$b$} (52);
		\end{tikzpicture}
		\caption{Iteration 2 \\ ($CC_{min}(\E, a,1,2)$)}
		\label{fig:iteration2}
	\end{subfigure}
	\begin{subfigure}[b]{.32\textwidth}
		\centering
		\begin{tikzpicture}[node distance={15mm}, thick, scale=1] 
			\node[ego, winnable] (13) {\Situation{1}{\none,\none,\none}}; 
			\node[alter, winnable, ,  inner sep=-1pt] (23) [right of=13] {\Situation{2}{0,\none,\none}}; 
			\node[ego, winnable] (73) [below right of=23] {\Situation{7}{0,\none,\none}}; 
			\node[ego, winnable] (33) [above right of=23] {\Situation{3}{0,\none,\none}}; 
			\node[alter, winnable, ,  inner sep=-1pt] (43) [above of=33] {\Situation{4}{0,0,\none}};
			\node[ego, winnable] (53) [left of=43] {\Situation{5}{0,0,\none}};  
			\node[alter, winnable,  inner sep=-1.5pt] (63) [below of=53] {\Situation{6}{1,0,0}};
			
			\draw[->] (-0.5,0.5) -- (13);
			\draw[->] (13) -- node[midway, above] {$\neg a$} (23); 
			\draw[->] (23) -- node[midway, below left] {$\neg b$} (73); 
			\draw[->] (23) -- node[midway, above left] {$b$} (33); 
			\draw[->] (33) -- node[midway, right] {$\neg a$} (43); 
			\draw[->] (43) -- node[midway, above] {$b$} (53); 
			\draw[->] (53) -- node[midway, right] {$a$} (63); 
		\end{tikzpicture}
		\caption{Iteration 3 \\ ($CC_{min}(\E, a,1,3)$)}
		\label{fig:iteration3}
	\end{subfigure}
	\caption{Situation graphs for the game in \cref{fig:smallGameIterationsMatter} with iteration over $CC_{min}(\E, a,1,7)$. More than three iterations are not necessary, since there already is a winning strategy for $\E$ in the third iteration.}
	\label{fig:situationGraphs}
\end{figure}

For illustrating the synthesis algorithm, we consider the game in \cref{fig:smallGameIterationsMatter} as small example. Circles represent locations controlled by $\E$. Diamond-shaped locations are controlled by $\A$. Let $CC_{min}(\E, a,1,7)$ be a counting constraint that $\E$ needs to satisfy. For the sake of keeping the example small, we pass on more counting constraints and only distinguish between the actions \enquote{$a$} and \enquote{$\neg a$} of $\E$. 
The constructed parts of the situation graphs for three iterations are shown in \cref{fig:situationGraphs}. Each state of the situation graph is marked with the respective state number of the game graph and with the history of last counting constraint-relevant turns of $\E$. The history length depends on the size of the counting constraint in the considered iteration. For example, the state marked with state 9 and history $(1,0)$ in \cref{fig:iteration2} encodes that $\E$ played $a$ in its last turn and played something else ($\neg a$) in its second to last turn. States highlighted with gray background are identified as being winnable.
The first iteration reduces the counting constraint to $CC_{min}(\E, a,1,1)$ (\enquote{$\E$ plays $a$ at least in one of 1 turns}), fully specifying how $\E$ is allowed to behave. The corresponding situation graph is shown in \cref{fig:iteration1}. State $2,(0)$ has no successor, since the counting constraint is already violated in this state. None of the states of the situation graph are in the winning region of the game. 
In the second iteration, the counting constraint for $\E$ is more relaxed, consequently the situation graph (\cref{fig:iteration2}) has more states. 10 of the states belong to the winning region of $\E$, since $\E$ can guarantee to avoid states with counting constraint violations (state $4,(0,0)$) from those states. Since the initial state is not marked as winnable, there exists no winning strategy for $\E$ and the third iteration is entered. 
In the situation graph for the third iteration (\cref{fig:iteration3}) the benefit of the iterated approach becomes visible: State $7,(0,-,-)$ is related to state $7,(0,-)$ of the previous iteration and since the latter one is already marked as winnable, so can state $7,(0,-,-)$. Hence, successors of $7,(0,-,-)$ do not need to be further considered. The same holds for state $6,(1,0,0)$, which is related to the winnable state $6,(1,0)$ of the second iteration. As a consequence, the situation graph of the third iteration is even smaller than the one of the second iteration. The initial state $1,(-,-,-)$ can now be marked as winnable, hence there already exists a winning strategy for $\E$ in the third iteration and no further iteration is required.
Please note that the focus on the example is to show how the situation graph evolves over multiple iterations, illustrating the benefit of the iterated approach. However, the example is too small to actually be significantly more efficient than synthesizing a winning strategy without iterations. 

A non-optimized explicit state implementation of \cref{alg:iteratedSynthesis} in Python was used to give an idea for the performance of iterated synthesis with counting constraints in larger examples. We will now sketch the insights retrieved from an exemplary game, solved with this implementation. The game graph had around 1.8mio states, 2.7mio transitions and a counting constraint for $\E$ of length 10. The algorithm took around 28 minutes to find a winning strategy for $\E$ in the $8^{th}$ iteration. Hence, iteration 9 and 10 were neither needed nor performed. The situation graph in the last required iteration had around 2.8mio states. For comparing those numbers with a non-iterated synthesis approach, \cref{alg:generateSituationGraph} and \cref{alg:findWinReg} were used to directly calculate a winning strategy for constraint length 8 for the same game. The calculation required around 4 times longer and used around 2.5 times more states. Note that limiting the constraint length directly to 8 was only possible because of the retrieved knowledge on strategy existence for this constraint length of the iterated synthesis calculations before. The comparison would even be more in favor of the iteration approach if the minimal counting constraint for which a winning strategy exists were not given for the non-iterative computation. We did not let the non-iterative algorithm run for the full counting constraint length of 10, since the expected amount of required states would have reached hardware limitations.
The realized comparison shows the great potential of successively enlarging counting constraints, allowing for incomplete graph constructions due to retrieved information on already winnable states of prior graphs instead of encoding the full constraints directly in a graph for strategy synthesis.

\section{Discussion and Future Work} \label{sec:discussion}

The exploitation of the monotony property inherent in counting constraints for iterative synthesis has demonstrated promising outcomes, indicating the potential for time- and memory-efficient computation of controllers for reactive systems. The current investigation aimed to explore the broader applicability of iterated synthesis utilizing counting constraints, an objective that has been achieved. However, certain challenges and considerations in the chosen game setting should be discussed in the following, paving the way for future research directions.

\begin{figure}
	\centering
	\begin{tikzpicture}[node distance={16mm}, thick, scale=0.8] 
		\node[ego] (1) {$1$}; 
		\node[alter] (2) [right of=1] {$2$}; 
		\node[ego] (3) [below of=2] {$3$};
		\node[ego] (4) [right of=2] {$4$};
		\node[alter] (5) [right of=4] {$5$};
		\node[ego] (6) [below of=5] {$6$};
		\node[alter] (7) [below of=4] {$7$};
		
		\draw[->] (-1,0) -- (1); 
		\draw[->] (1) -- node[midway, above] {$a$} (2); 
		\draw[->] (2) to [out=225,in=135]  node[midway, left] {$b$} (3); 
		\draw[->] (3) to [out=45,in=315]  node[midway, right] {$a$} (2); 
		\draw[->] (2) -- node[midway, above] {$\neg b$} (4); 
		\draw[->] (4) -- node[midway, above] {$a$} (5); 
		\draw[->] (5) -- node[midway, right] {$\neg b$} (6); 
		\draw[->] (6) -- node[midway, above] {$a$} (7); 
		\draw[->] (7) -- node[midway, left] {$b$} (4); 
	\end{tikzpicture} 
	\caption{$\A$ can always fulfill the counting constraint $CC_{min}(\A, b, 1, 3)$, but can run into a violation for $CC_{min}(\A, b, 1, 2)$.}
	\label{fig:invalidGame}
\end{figure}
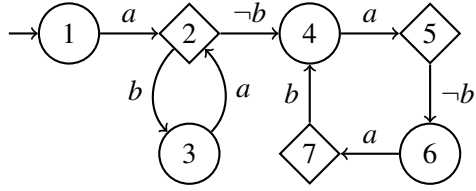

\textbf{Towards cooperative games}: As already mentioned above, the idea of adding window counting constraints like \enquote{The player $\A$ plays $a$ at least (or: at most) $k$ times out of $l$ of its own turns.} for the other player $\A$ seems obvious. In the current setting, we apply the synthesis algorithm on the winning region of the underlying safety game. If $\A$-constraints are added, the previous winning region (without counting constraints) would only be an under-approximation of the winning region for the safety game with counting constraints. Hence, the synthesis algorithm may fail to find an existing winning strategy for $\E$. The problem can be solved by omitting the calculation of the winning region beforehand and integrate the safety condition in the iterated synthesis approach. This can be done by handling unsafe states the same way as states in which $\E$ violates its constraints. If we want to stay in a zero-sum game setting, we could restrict the games of interest to those in which $\A$ can actually fulfill its constraints. The following property could be added to the definition of a game with counting constraints (\cref{def:gameWithCounstraints}). $\A$ cannot be forced into constraint violations: For each prefix $\pi(n) = \pi_0 \sigma_0 \pi_1 \dots \pi{n-1} \sigma_{n-1} \pi_n$, $n \in 2\N+1$, of a play on $G$ that satisfies all counting constraints of $\A$, there exists $(\pi_n, a , \pi_{n+1})\in \rightarrow$, such that $\pi_0 \sigma_0 \pi_1 \dots \pi_{n-1} \sigma_{n-1} \pi_n a \pi_{n+1}$ is also a prefix of a play on $G$ that satisfies the counting constraints of $\A$. 
This property simplifies the formulation of winning conditions for $\E$, circumventing complex scenarios arising from ambiguous outcomes wherein one player forces the other into constraint violations at the expense of own future constraint violations.
However, this restriction is limiting the possibility to iterate over counting constraints to constraints of $\E$. In general, a game with counting constraints may satisfy the requirement of $\A$ always being able to adhere to its counting constraints, only to find the requirement violated for the game with a modified counting constraint as used in the iterations. An illustrative example is provided in \cref{fig:invalidGame}. $\A$ has the counting constraint  $CC_{min}(\A, b, 1, 3)$, i.e.\ $\A$ plays $b$ at least once in three of its turns. Recall that \enquote{$\A$ cannot be forced into constraint violations} is defined in \cref{def:gameWithCounstraints}  as $\A$ is able to enlarge each prefix that satisfies the constraint such that the resulting prefix is also satisfying the constraint. This property is fulfilled when considering the game graph and the constraints $CC_{min}(\A, b, 1, 1)$ or $CC_{min}(\A, b, 1, 3)$. However, it is violated for $CC_{min}(\A, b, 2, 3)$, since the prefix $(1, a, 2, \neg b, 4, a, 5)$ satisfies the constraint\footnote{$\A$ could play $b$ forever to complete the prefix to an infinite play that satisfies the constraint. This play is not in $G$, but nonetheless is sufficient according to the definition of a prefix satisfying  a constraint in \cref{def:windowCountingConstraints}.}, but there is no possibility for $\A$ to still satisfy the constraint with the next turn. Since the definition of a winning strategy relies on the game property of $\A$ not be forceable into counting constraint violations, \cref{theorem:monotony} cannot be extended to iterations over $\A$-constraints. However, such an extension would offer additional potential for more efficient synthesis algorithms.

We plan to approach this problem by leaving the zero-sum setting. The environment wins if it has a strategy that guarantees to satisfy all of its counting constraints. In particular, it is possible that the environment violates a constraint and loses. The envisioned game setting shall avoid the well-known problem of $\E$ winning only by falsifying the assumptions in form of counting constraints on $\A$. Instead, $\E$ shall support $\A$ in satisfying all constraints as long as this does not compromise the adherence of own constraints. This leads us in the direction of searching for strategy profiles with certain properties as synthesis results instead of winning strategies only for $\E$ with the exact profile properties yet to be determined. It can be foreseen that this setting requires more synchronization between the players than that presented by a zero-sum setting, in which $\A$ did not even need knowledge on counting constraints of $\E$.

\textbf{Extension of counting constraint types:} It is worth to consider additional specification patterns with similar monotony properties as the presented counting constraints. For instance, a pattern like \enquote{if $x$ is played, $\E$ plays $y$ after at most $k$ turns} is frequently used as specification. Satisfying such a specification becomes easier for larger $k$. In terms of an iterative algorithm: states of the situation graph for some iteration are winnable, if the related state is winnable in an earlier iteration. The identification of additional counting constraints and the adaption of the iterative strategy synthesis algorithm to such constraints increases the applicability of the approach to more systems.

\textbf{Combination of various counting constraints:} In the presented synthesis algorithm, iteration is only done over one counting constraint. All other constraints remain fixed. We anticipate greater savings in memory and computational time than already provided by the presented algorithm by iterating over several constraints (successively or alternating). Such an extension is expected to require only manageable modifications of the existing algorithm for sets of counting constraints that use the same type of information from one iteration to the other (e.g.\ exclusively on winnable states of the various situation graphs). The iteration over a set of constraints that use different types of information during iteration (e.g.\ on winnable states for some of the constraints and on non-winnable states for other constraints) is expected to require a more thorough adaption of the algorithm.

\textbf{Symbolic representation:} The presented synthesis approach uses an explicit representation of states in the situation graph as arena. However, symbolic synthesis showed to be significantly more efficient than explicit synthesis algorithms for many (but not all) applications \cite{Finkbeiner2016}. Since the presented approach already has similarities to antichains and the states of the arena have a special structure (representing a snippet of the history of a play), we expect that the approach can be transformed in a symbolic algorithm. We plan to investigate a symbolic version of the algorithm and to compare its performance with its explicit version.

\section{Conclusion}\label{sec:conclusion}
Synthesis algorithms for reactive systems are promising tools for various engineering tasks, most prominently for the creation of correct-by-construction controllers and for checking the feasibility of specifications. The efficiency of such algorithms is a challenge for getting synthesis into application, since the translation of the system specification into an automaton that is suitable for synthesis is costly in terms of memory and computational time. 
The exploitation of specific properties in the specification can help to overcome this challenge. 
In this paper, we have shown the potential of iterative synthesis algorithms for specifications with monotony properties as for the presented counting constraints. 
With each iteration, the automaton encoding the specification is becoming larger.
The key idea is to gather information in each iteration that can be used in the next iteration to reduce the size of the automaton. The precise nature of this information depends on the considered specification. We have shown an iterated algorithm for a counting constraint of the form \enquote{the system does a specific move $m$ at least in $k$ turns out of $l$}, in which information on winnable states of the automaton in one iteration can be used to determine which parts of the automaton for the next iteration do not need to be constructed. In the presented example, the iterative approach requires significantly less memory and computational time than direct synthesis with full specification translation into one automaton. 
As future work, we plan to extend the iterative synthesis in four dimensions: (1) Consideration of more cooperative behavior between system and its environment instead of a purely adversarial setting, (2) identification of new specification types with monotony properties that can be exploited via iterated synthesis, (3) development of algorithms that use advantages of different specification types simultaneously and (4) transformation of the synthesis approach to a symbolic synthesis algorithm.

\bibliographystyle{eptcs}
\bibliography{references}
\end{document}